\documentclass[journal]{IEEEtran}
\usepackage{amssymb}
\usepackage{amsmath}
\usepackage{enumerate}
\usepackage{multirow}
\usepackage{subfigure}
\usepackage{graphicx}
\usepackage{url}
\usepackage[flushleft]{threeparttable}
\usepackage{extarrows}
\usepackage{algorithm,color}

\usepackage{algpseudocode}

\ifCLASSINFOpdf
\else
\fi
\begin{document}
%
\title{Accelerating Secure and Verifiable Data Deletion in Cloud Storage via SGX and Blockchain}



\author{Xiangman Li and Jianbing Ni \thanks{X. Li and J. Ni are with the Department of Electrical and Computer Engineering and Ingenuity Labs Research Institute, Queen's University, Kingston, Ontario, Canada K7L 3N6. Email: jianbing.ni@queensu.ca.}   }

\maketitle

\begin{abstract}

Secure data deletion enables data owners to fully control the erasure of their data stored on local or cloud data centers and is essential for preventing data leakage, especially for cloud storage. However, traditional data deletion based on unlinking, overwriting, and cryptographic key management either ineffectiveness in cloud storage or rely on unpractical assumption. In this paper, we present SevDel, a secure and verifiable data deletion scheme, which leverages the zero-knowledge proof to achieve the verification of the encryption of the outsourced data without retrieving the ciphertexts, while the deletion of the encryption keys are guaranteed based on Intel SGX. SevDel implements secure interfaces to perform data encryption and decryption for secure cloud storage. It also utilizes smart contract to enforce the operations of the cloud service provider to follow service level agreements with data owners and the penalty over the service provider, who discloses the cloud data on its servers. Evaluation on real-world workload demonstrates that SevDel achieves efficient data deletion verification and maintain high bandwidth savings.

\end{abstract}


%
\IEEEpeerreviewmaketitle

\section{Introduction}
Outsourcing data to the cloud storage is a common practice for data owners to save the burden of self-managing massive data \cite{abu2010racs}. The data owners can on-demand rent the storage spaces provided by the cloud service providers. The outsourcing data management enables the data owners to access their data at anytime and from anywhere. Due to the appealing features, the cloud storage services, such as Amazon S3 \cite{palankar2008amazon}, Google Drive \cite{quick2014google}, Dropbox \cite{dropbox2014dropbox}, Apple iCloud \cite{holt2018apple}, and Microsoft OneDrive \cite{wilson2015onedrive}, have attracted a large number of stable and loyal users. Data security is one of the primary concerns for data owners. After data owners outsource their data to the cloud data centers, they lose their physical control over their data. Thus, the data owners have no choice, but relying on the cloud service providers to protect their data. Unfortunately, due to the frequently happened data leakage or breach accidents, security are always ranked as the top threats in cloud storage \cite{kuyoro2011cloud}, although the cloud service providers have great efforts for guaranteeing the confidentiality, integrity, and availability of the outsourced data on their cloud servers.

Data deletion \cite{zheng2020toward}, as one of important security technologies, has not received sufficient attentions from both data owners or cloud service providers in cloud storage. It provides methods to securely erase data from local storage medium or remote cloud servers, which can significantly reduce the probability of data leakage. Moreover, privacy regulations, such as GDPR \cite{voigt2017eu}, CPPA \cite{hughes2021symposium}, and PIPL \cite{calzada2022citizens}, have clearly defined the principles of data deletion, called right to be forgotten. The data centers should delete the personal data if they are no longer necessary to the purpose for which it was collected, and the data owners have the right to request the data centers to delete their personal data stored on the data centers. Therefore, data deletion becomes increasingly critical for the service providers, which should provide effective ways to guarantee secure data deletion.

\subsection{Related Work}
It is not a trivial problem to securely delete data. It is well recognized that there is no existing software-based solution that can provide complete data removal from storage medium. Existing
deletion methods can be summarized in the following categories:

Deletion by unlinking. This method is widely deployed on the file management system in operating systems, such as Windows, IOS, and Linux. When the user would like to delete a file (i.e., press the "delete" button), the operating system delete the link of the file from the file systems, and returns "success" to the user. The file is no longer accessible because the link of the file is removed. Nevertheless, this is not the real file deletion, as the file content still remains on the disk. An adversary can simply use a file recovery tool to access the deleted file by scanning the disk \cite{garfinkel2003remembrance}.

Deletion by block erasure. This method is utilized by storage mediym, such as solid-state drive (SSD) to securely clean the data. It applies a voltage spike to all available flash memory blocks in unison. Each block is altered with a vendor-specific value and SSD become "clean" \cite{liu2017erasucrypto}. However, this method erases all the data on the drive and does cause a small amount of wear.

Deletion by overwriting. Overwriting is an important tool to delete the data by overwriting the data with new, insensitive data, e.g., all zeros. There are multiple tools that can offer 35-pass overwriting times. However, one inherent limitation with the overwriting methods is that they cannot guarantee the complete removal of data. It is effectively impossible to sanitize storage locations by simply overwriting them, no matter how many overwrite passes are made
or what data patterns are written \cite{gutmann1996secure}. The conclusion holds for not only magnetic drives, but also tapes, optical disks, and flash-based solid state drives. In all these cases, an attacker, equipped with
advanced microsoping tools, may recover overwritten data based on the physical remanence of the deleted
data left on the storage medium. Therefore, although overwriting data makes the recovery harder, it does
not change the basic one-bit-return protocol.

Deletion by encryption. Boneh and Lipton \cite{boneh1996revocable} proposed the first cryptography-based method for secure date deletion by encrypting data before saving it to the disk, and deleting the data by discarding the decryption key after encryption. This method is desirable when duplicate copies of data are backed up in distributed. However, this method essentially change the problem of deleting a large amount of data to the problem of deleting a short key. However, forgetting a decryption key is non-trivial. The key can be stored on a hard disk is not easy to be permanently deleted, i.e., never be recoverable for an adversary even it obtain the storage medium \cite{hao2015deleting}.

However, the problem of key deletion becomes dramatically difficult as the cloud server performs the encryption of the outsourced data in traditional secure cloud storage. Although some cloud storage services enable user-side encryption, i.e., the data owners also can encrypt their data before outsourcing, the server-side encryption is more general. The cloud server encrypts the data after receiving them from the data owners with an encryption key and decrypts the data that the owners would like to access before returning them to the data owners. In this model, the encryption is fully controlled by the cloud servers, which brings the worries of the data owners about the encryption of their outsourced data and secure deletion of the decryption keys.

\subsection{Contributions}
In this paper, we propose a novel secure and verifiable data deletion scheme, named SevDel, for cloud storage. To reduce the concern that whether the cloud server honestly encrypts the outsourced, we utilize the randomly sampling method and the zero-knowledge proof \cite{goldreich1996composition} to verify the encryption without retrieving the ciphertexts of the outsourced data. The encryption is also performed based on the Intel SGX \cite{gotzfried2017cache} to prevent the possible data leakage. The enclave is created for each file for the encryption and the management of the keys. Thus, the operation of the deletion of the decryption becomes the destroy of the enclave. In addition, to enforce the cloud servers to protect the outsourced data, the smart contract is designed based on the service-level agreements between the data owners and the cloud service providers. We demonstrate the properties of confidentiality, verifiability, erasability, and auditability of SevDel through security analysis and show that the proposed SevDel has outstanding performance for deployment.

%

\section{System and Security Models}\label{sec:model}
In this section, we introduce the system model and security model of our SevDel.

\subsection{System Model}
We present the system model of SevDel, that comprises three kinds of entities: 1) a data owner that outsources the data to the cloud and requests to delete them after the data is processed or used; 2) a cloud service provider that offers secure cloud storage services (i.e., the outsourced data of data owners are encrypted by the service provider with its chosen secret keys or by the data owners before outsourcing) to data owners with its storage servers in the cloud data center, and each server has high-performance hard disks for data storage and has the Intel Core that supports for SGX \cite{gotzfried2017cache}; and 3) a blockchain node \cite{zheng2018blockchain} that participates the blockchain network to maintain transactions happened between two parties. The blockchain can be the public blockchain, e.g., Bitcoin blockchain, Ethereum blockchain, or Hyperledger. It maintains an automatically executable smart contract that enforces the penalty on the cloud service provider if it leaks the outsourced data of users.

Intel SGX \cite{gotzfried2017cache}, a suite of security-related instructions built into modern Intel CPUs, can create a hardware-protected environment, enclave, for shielding the execution of code and data. An enclave resides in a hardware-guarded memory region called the enclave page cache (EPC) for hosting any protected code and data.
In enclave, SGX performs the encryption of the outsourced data with a secret key stored on the EPC. The deletion of the encrypted data for the data owner is the deletion of the secret key in enclave. More specifically, the secret in enclave is erased after the enclave is destroyed.

\subsection{Threat Model}
The security threats are mainly from the outsider attackers or the data thief. An outsider attacker or a data thief may compromise the cloud server to steal the data on the hard disks. The frequently happened data leakage incidents on cloud have witnessed the risks of cloud storage services. This risk is high because of potential code vulnerability, and the damage is severe as the data leakage incidents significantly affect reputation. Moreover, the employees in cloud service may steal the data on cloud servers. We have witnessed many data corruption or leakage incidents that occur due to the operation errors or misbehavior of the employees. The main security objective is to protect the cloud data for users against data leakage incidents.

A cloud service provider is the legitimate processor of the Intel SGX and holds the service level agreements with the data owners for maintaining outsourced data. It is expected that the cloud service provider stores the encrypted outsourced data of data owners on the hard disks of cloud servers and deletes the data under the requests of data owners or based on the principles of privacy regulations, like GDPR and CPPA, and PIPEDA. It is assumed that the cloud service provider may not deviate from the expectation due to the agreement with data owners, that is, the cloud service provider is rational. It follows the service level agreements to honestly offer data storage services. Undoubtedly, regulating the implementation of the agreement between the users and the cloud service provider become necessary.

A data owner is an honest party to rent storage spaces from the cloud storage services and outsources the data to the cloud servers in the data center. The data owner chooses the reliable service providers for data outsourcing. According to the modes for protecting cloud data in cloud storage, e.g., Amazon S3 of Amazon Web Service, the owners can determine whether to encrypt their data before outsourcing. The data owners can use secret keys to encrypt their data before outsourcing. If the owners do not encrypt the data, the cloud server chooses the secret keys for data encryption. In this paper, we study secure data erasure for the latter case because it is trivial to achieve data deletion if the data owners encrypt their data by themselves, as they can delete their keys and then no one can read the cloud data.

\section{Proposed SevDel}
In this section, we propose the overview and the detailed construction of our SevDel.

\section{Overview}
Our SevDel accelerates security and verifiability of cloud data erasure in cloud storage. It can serve as the central element of secure cloud storage and erasure in cloud storage services, such as Amazon S3 Find and Forget, the solution to selectively erase records from data lakes stored on Amazon S3. To prevent data leakage, the received file from the data owner is encrypted by the cloud server with a randomly selected private key with additive homomorphic encryption, such as lifted ElGamal encryption \cite{ni2017differentially}. The encryption operation is performed in the enclave of Intel SGX. The encryption of the file is audited by the data owner to ensure that the file is correctly encrypted as claimed by the cloud service provider. The random sampling is utilized to enable probabilistic auditing of the encrypted data and the ciphertexts are aggregated to compress auditing messages. The cloud server proves to the data owner that the entire file is encrypted with lifted ElGamal encryption by a randomly chosen key with a large probability, without retrieving the encrypted file. The challenge here is to ensure that the proved ciphertext is really the encryption of the correct outsourced file. To blind the file and its ciphertext during auditing, the cloud server should prove that the plaintext of the ciphertext is the outsourced file in the homomorphic authentication tags, which are produced by the data owner and outsourced along with the file. Meanwhile, they can also used to verify the integrity of the outsourced file based on provable data possession \cite{erway2015dynamic} or proof of retrievability \cite{shacham2013compact}.

The deletion of the oursourced file on the cloud server is enabled by the deletion of the secret key of the file. If the secret key is permanently deleted, no one is able to decrypt the ciphertext. The secret key deletion is realized by the Intel SGX. An enclave is created when the cloud server receives the file and the encryption is performed in the enclave. Also, the encryption key is stored in the enclave. In order to permanently forget the key, the simple way is to destroy the corresponding enclave.

To ensure the cloud service provider to honestly maintain and encrypt outsourced files of data owners, a smart contract is created based on the service level agreement between the service provider and data owners. the deposits of the service provider are made when the cloud storage service is bootstrapped. The deposits are paid to the data owner if the file of the data owner is found on the Internet, which means that the file is leaked during storage. The condition to trigger the payment is the key point of the smart contract. We convert this data leakage problem to be the provable data possession. If a data owner is succeed to giving a proof that she possesses the encrypted version of her outsourced files, the penalty is performed over the service provider and a certain amount of the deposits is transferred to the data owner. The conversion is valid because only the cloud server has the encrypted version of the outsourced file of the data owner. The cloud server performs encryption after receiving the outsourced file and decryption before returning it to the data owner. The ciphertext of the outsourced file should be only known by the cloud server. Although the data owner knows the cleartext of the file, the data owner obtains the same ciphertext, as the data encryption on the side of the cloud server is probabilistic.

Our SecDel consists of the following algorithms.

{\sf Setup}: This algorithm is run by the cloud service provider to bootstrap the cloud storage systems. With the input of the security parameter, the algorithm outputs the system parameters and the public-private key pairs of the cloud servers.

{\sf Contract}: This algorithm is run by the cloud service provider to initialize a smart contract that implements the service level agreement with the data owners. The smart contract is maintained by the blockchain nodes.

{\sf KeyGen}: This algorithm is run by the data owner. With the input of the system parameters, the algorithm takes the input of the system parameter and generate the public-private key pair of the data owner for data outsourcing.

{\sf Outsource}: This algorithm is run by the data owner to outsource the file to the cloud server. With the input of the security parameters, the private key of the data owner, and the outsourcing file, the algorithm produces the homomorphic authentication tags of the data blocks of the file and outsource the file, along with the generated tags.

{\sf Encrypt}: This algorithm is run by the cloud server that encrypts the received file with a randomly chosen private key. With the input of the file, the private key of the cloud server, and the chosen private key, the algorithm outputs the encrypted file, the corresponding public key, and the homomorphic authentication tags of the data blocks of the encrypted file.

{\sf Verify}: This is an interactive protocol between the cloud server and the data owner to audit the encryption of the outsourced file. The data owner randomly samples the data blocks, and the cloud server generates a proof that proves the encryption of the sampled data blocks. The data owner finally verifies the proof to learn whether the file has been encrypted by the cloud server.

{\sf Delete}: This algorithm is run by the cloud server who deletes the file under the request of the data owner or the data is no longer needed for data analysis.

{\sf Audit}: This is an interactive protocol between the data owner and the blockchain nodes. The blockchain nodes randomly samples the data blocks owned by the data owners and the data owner responds the proof that proves the ownership of the encrypted data block. Then, the blockchain nodes verify the proof to learn whether the file has been disclosed. If the proof is valid, the smart contract is executed to give penalty to the cloud service provider.

The correctness of SevDel has the following aspects: 1) The encryption of the outsourced file should be correctly recovered by the cloud server with the corresponding secret key; 2) the data owner can identify that the cloud server does not encrypt the oursourced file on hard disks as agreed with the service level agreement; 3) the deleted outsourced file can be no longer recovered; and 4) the blockchain node can execute the penalty if the data owners find the leaked outsourced data.

\subsection{Detailed SevDel}
{\sf Setup}: Let $q$ be a large prime and $\mathbb G_1$, $\mathbb G_2$ and $\mathbb G_T$ be three multiplicative cyclic groups of the same prime order $p$. $g_1$ and $g_2$ are the generators of~$\mathbb G_1$ and $\mathbb G_2$, respectively. ${e}:\mathbb G_1 \times \mathbb G_2 \rightarrow \mathbb G_T$ denotes an admissible bilinear pairing.

The file $M$ to be outsourced is divided into $n$ blocks and each block is further split into $s$ sectors. Thus, the fiel is denoted as ${M}={\{m_{ij}\}}_{i \in [1,n],j \in [1,s]}$ and the abstract information of $M$ is denoted as ${\mathbb I}_{M}$.
$H:\{0,1 \}^* \rightarrow {\mathbb G_1}$ is a cryptographic hash function that maps the ${\mathbb I}_{M}$ to a point in $\mathbb G_1$.

The cloud service provider chooses a random number $a \in
{\mathbb Z}_p $ and calculates $A=g_2^{a} \in {\mathbb G_2}$. The private key of the data owner is $a$, and the corresponding public key is $A$.

{\sf Contract}: The service provider creates the smart contract CS-SevDel to provide cloud storage services to data owners. To provide the service, the service provider initiates CS-SevDel.Init to setup the smart contract and deposits an amount of money on the blockchain as insurance in CS-SevDel.Service. The a part of the deposit would be sent to the data owner if the outsourced data is leaked and the remainder would be re-fund to the service provider.

{\sf KeyGen}: An data owner chooses a random number $w \in
{\mathbb Z}_p $ and calculates $W=g_2^{w} \in {\mathbb G_2}$. The private key of the data owner is $w$, and the corresponding public key is $W$.

{\sf Outsource}: The data owner chooses $s$ random values $x_1,\cdots, x_s \in {\mathbb Z}_p $ and
computes $u_j={g_1}^{x_j} \in {\mathbb G_1}$ for $j \in [1,s]$.
Then, for each block $m_i$ ($i \in [1,n]$), it computes a tag $t_i$ as
\begin{center}
$\phi_i=(H({\mathbb I}_{M}||i)\cdot \prod_{j=1}^{s} {u_j}^{m_{ij}})^{w}$.
\end{center}
The data owner outputs the set of homomorphic authentication tags $T={\{\phi_i\}}_{i \in [1,n]}$. The tag set $\Phi$, the file index ${\mathbb I}_{M}$, and the file $M$ are sent to the cloud server.

{\sf Encrypt}: After receiving $({\mathbb I}_{M},M,\Phi)$ from a data owner, the cloud server first randomly selects a private key $v \in {\mathbb Z}_p$ and computes $V=g_1^v \in {\mathbb G_1}$. The cloud server uses the random private key $v$ to encrypt each data block of the received file $m_{ij}$ as $E_{ij}=(E'_{ij},E''_{ij})=(g_1^{m_{ij}}V^{r_{ij}}, g_1^{r_{ij}})$, where $r_{ij}$ is a random number chosen from ${\mathbb Z}_p$. The set of the encrypted blocks is denoted as $E={\{E_i\}}_{i \in [1,n]}$. Then, for each encrypted block $E_i$ ($i \in [1,n]$), the cloud server computes a homomorphic authentication tag $\sigma_i$ for the encrypted block as
\begin{center}
$\sigma_i=(H({\mathbb I}_{M}||i)\cdot \prod_{j=1}^{s} {u_j}^{E'_{ij}}{v_j}^{E''_{ij}})^{a}$.
\end{center}
The set of the tags of the encrypted blocks is denoted as $\Sigma={\{\sigma_i\}}_{i \in [1,n]}$. Finally, the cloud server stores $({\mathbb I}_{M},E, \Sigma)$ on the hard disks and uploads $({\mathbb I}_{M}, \Sigma)$ to the blockchain.

{\sf Verify}: To verify the encryption of the outsourced file $M$, the data owner takes the abstract information ${\mathbb I}_{M}$ as inputs. It selects some data blocks to construct a \emph{challenge set $Q$} and picks a random $l_i \in {\mathbb Z}_p^*$ for each $m_i$ $(i \in Q)$. The challenge $(i, l_i)_{i \in Q}$ is sent to the cloud server.

To respond the challenge, the cloud server generates $P_1$ as
\begin{center}
$P_1=\prod_{i \in Q} g_1^{l_im_{ij}}V^{l_ir_{ij}}.$
\end{center}

The cloud server computes $Q_j=\sum_{i \in Q} {l_i}\cdot {m_{ij}}$ for each $j \in [1,s]$.
Then, it computes $Q_2$ as
\begin{center}
$P_2=\prod_{j=1}^s \phi_i^{l_i}.$

$\pi \leftarrow NIZK \{(Q_j, r_{ij}): P_1=\prod_{i \in Q} g_1^{l_im_{ij}}V^{l_ir_{ij}}, P_2=\prod_{j=1}^s \phi_i^{l_i}\}$.
\end{center}
The data owner verifies the validity of the zero-knowledge proof $\pi$ to determine whether the outsourced file has been encrypted or not.

{\sf Delete}: The cloud server deletes the random private key $v$ that is used to encrypt the file $F$ by destroying the enclave that used to store $v$. The cloud server creates an enclave for each file received and use the enclave to maintain the private key.

{\sf Audit}: If the data owner obtains the leaked encrypted file $F$, the data owner can prove to the blockchain nodes that the cloud server has data leakage. The blockchain node selects some data blocks to construct a \emph{challenge set $R$} and picks a random $\gamma_i \in {\mathbb Z}_p^*$ for each $E_i$ $(i \in R)$. The challenge $(i, \gamma_i)_{i \in R}$ is sent to the data owner.

To respond the challenge, the data owner generates $Q_1$ as
\begin{center}
$Q_1=\prod_{i \in R} \gamma_i E_{ij}.$
\end{center}

Then, it computes $Q_2$ as
\begin{center}
$Q_2=\prod_{j=1}^s \sigma_i^{\gamma_i}.$
 \end{center}
The data owner returns $(Q_1, Q_2)$ to the blockchain node. The blockchain node verifies $(Q_1, Q_2)$ to determine whether the cloud server has disclosed the file $F$. If yes, the blockchain node performs CS-SevDel.Penalty to give penalty to the cloud service provider.

The correctness of SevDel can be check that 1) the encryption of the outsourced file is correctly recovered; 2) the verification equation can pass; 3) the security of Intel SGX; and 4) the blockchain node can execute the penalty.

\section{Security of SevDel}
The security of SevDel should capture the properties of confidentiality, verifiability, erasability, and auditability.

The confidentiality of the outsourced file relies on the semantic security of the data encryption scheme used by the cloud server. SevDel utilizes the lifted ElGamal encryption scheme to encrypt each block of the outsourced file $M$. Here, each block is independently encrypted with the key $V$. As the lifted ElGamal encryption scheme can be proved semantic security under the Decisional Diffie-Hellman (DDH) assumption, the confidentiality of the outsourced file is achieve as long as the DDH assumption holds.

\begin{picture}(-10,608)
\put(0,10){\framebox(248,600)}
\put(0,554){\makebox(100,100)[l]{~~~~~~~~~~~~~~~~~~Smart Contract {CS-SevDel}}}
\put(0,542){\makebox(200,95)[l]{~~~~~\textbf{Init}: Set {state:=INIT}, {File}$:=\{\}$, {Onwer}$:=\{\}$, {RU}$:=\{\}$,}}
\put(0,530){\makebox(200,95)[l]{~~~~~~~~~~~{{Tags}$:=\{\}$, { Param:=SerDel}$(1^{\lambda})$}.}}
\put(0,518){\makebox(200,95)[l]{~\textbf{Service}: Upon receiving (``Create", $N$, $file$, $A$, Deposit,}}
\put(0,506){\makebox(200,95)[l]{~~~~~~~~~~ $T_1,T_2,T_3,T_4$) from a service provider $\mathcal{S}$:}}
\put(0,494){\makebox(200,95)[l]{~~~~~~~~~~~~~~~Assert {state=INT}.}}
\put(0,482){\makebox(200,95)[l]{~~~~~~~~~~~~~~~Assert current time $T\leq T_1$.}}
\put(0,470){\makebox(200,95)[l]{~~~~~~~~~~~~~~~Assert {ledger}$\mid\mathcal{S}\mid\geq$ \$Deposit.}}
\put(0,458){\makebox(200,95)[l]{~~~~~~~~~~~~~~~{ledger} $\mid\mathcal{S}\mid$:={ledger}$\mid\mathcal{S}\mid$--\$Deposit.}}
\put(0,446){\makebox(200,95)[l]{~~~~~~~~~~~~~~~Set {state:=CREATED}.}}
\put(0,434){\makebox(200,95)[l]{~~~~~~~~~~~~~~~Set {Accept}:=0.}}
\put(0,422){\makebox(200,95)[l]{~~~~~~~~~~~~~~~{File}:={File}$\cup$\{$\mathcal{S},N$,$A$,Deposit,{Accept},$T_{j=1-4}$\}.}}
\put(0,410){\makebox(200,95)[l]{~\textbf{Agree}: Upon receiving (``Accept", $\mathcal{U}_i,N,R_i$) from a}}
\put(0,398){\makebox(200,95)[l]{~~~~~~~~~~ data owner $\mathcal{U}_i$:}}
\put(0,386){\makebox(200,95)[l]{~~~~~~~~~~~~~~~Assert {state=CREATED}.}}
\put(0,374){\makebox(200,95)[l]{~~~~~~~~~~~~~~~Assert $T_1 \leq T \leq T_2$.}}
\put(0,362){\makebox(200,95)[l]{~~~~~~~~~~~~~~~Assert \$R$_i >$0.}}
\put(0,350){\makebox(200,95)[l]{~~~~~~~~~~~~~~~Assert {ledger}$\mid\mathcal{U}_i\mid\geq$ \$R$_i$.}}
\put(0,338){\makebox(200,95)[l]{~~~~~~~~~~~~~~~{ledger} $\mid\mathcal{U}_i\mid$:={ledger}$\mid\mathcal{U}_i\mid$--\$R$_i$.}}
\put(0,326){\makebox(200,95)[l]{~~~~~~~~~~~~~~~Set {Accept}:={Accept}+1.}}
\put(0,314){\makebox(200,95)[l]{~~~~~~~~~~~~~~~Set {state}$_i$:={ACCEPTED}.}}
\put(0,302){\makebox(200,95)[l]{~~~~~~~~~~~~~~~{Owner}$_N$:={Owner}$_N\cup\{\mathcal{U}_i$\}.}}
\put(0,290){\makebox(200,95)[l]{~~\textbf{Claim}: Current time $T=T_2$:}}
\put(0,278){\makebox(200,95)[l]{~~~~~~~~~~~~~~~Assert {state}$_i$={ACCEPTED}.}}
\put(0,266){\makebox(200,95)[l]{~~~~~~~~~~~~~~~Assert the data outsourcing $N$.}}
\put(0,254){\makebox(200,95)[l]{~~~~~~~~~~~~~~~Set {state:=CLAIMED}.}}
\put(0,242){\makebox(200,95)[l]{~\textbf{Audit}: Upon receiving (``Audit", $\mathcal{U}_i,N,c_i,d_i,\sigma_i,e_i,rk_i,$}}
\put(0,230){\makebox(200,95)[l]{~~~~~~~~~~~~$\mathcal{PK}_i$) from $\mathcal{U}_i$:}}
\put(0,218){\makebox(200,95)[l]{~~~~~~~~~~~~~~~Assert {state=CLAIMED}.}}
\put(0,206){\makebox(200,95)[l]{~~~~~~~~~~~~~~~Assert $T_{2} \leq T\leq T_{3}$.}}
\put(0,194){\makebox(200,95)[l]{~~~~~~~~~~~~~~~Assert $\mathcal{U}_i\in${AU}$_N$.}}
\put(0,182){\makebox(200,95)[l]{~~~~~~~~~~~~~~~Assert $\mathcal{PK}_i=1$.}}
\put(0,170){\makebox(200,95)[l]{~~~~~~~~~~~~~~~Set {state}$_i$:={UPLOADED}.}}
\put(0,158){\makebox(200,95)[l]{~~~~~~~~~~~~~~~Set {ledger} $\mid\mathcal{U}_i\mid$:={ledger}$\mid\mathcal{U}_i\mid$+\$R$_i$.}}
\put(0,146){\makebox(200,95)[l]{~~~~~~~~~~~~~~~{Owner}$_N$:={owner}$_N\cup\{\mathcal{U}_i$\}.}}
\put(0,134){\makebox(200,95)[l]{~~~~~~~~~~~~~~~{File}$_N$:={File}$_N\cup\{(\mathcal{U}_i,N,\sigma_i,e_i,rk_i)$\}.}}
\put(0,122){\makebox(200,95)[l]{~\textbf{Refund}: $T_{3} \leq T\leq T_{4}$ and {Owner}$_N$={File}$_N$:}}
\put(0,110){\makebox(200,95)[l]{~~~~~~~~~~~~~~~Set {state:=FULFILLED}.}}
\put(0,98){\makebox(200,95)[l]{~~~~~~~~~~~~~~~Set {ledger} $\mid\mathcal{U}_i\mid:=${ledger}$\mid\mathcal{U}_i\mid$+\$Deposit$_i$.}}
\put(0,86){\makebox(200,95)[l]{~~~~~~~~~~~~~~~Assert \$Deposit=$\sum_{i=1}^n$\$Deposit$_i$.}}
\put(0,74){\makebox(200,95)[l]{~~~~~~~~~~~~~~~Set {state:=FINISHED}.}}
\put(0,62){\makebox(200,95)[l]{~\textbf{Penalty}: $T_{3} \leq T\leq T_{4}$ and {AU}$_N\supset${RU}$_N$:}}
\put(0,50){\makebox(200,95)[l]{~~~~~~~~~~~~~~~Set {state:=UNFULFILLED}.}}
\put(0,38){\makebox(200,95)[l]{~~~~~~~~~~~~~~~{ledger}$\mid\mathcal{U}_i\mid:=${ledger}$\mid\mathcal{U}_i\mid$+\$R$^*_i$, for $\mathcal{U}_i \in ${RU}$_N$.}}
\put(0,21){\makebox(200,95)[l]{~~~~~~~~~~~~~~~Assert $\sum\limits_{i\in\{{\texttt{AU}}_N-{\texttt{RU}}_N\}}$\$R$_i$ =$\sum\limits_{i\in\{{\texttt{RU}}_N\}}$\$R$^*_i$.}}
\put(0,6){\makebox(200,95)[l]{~~~~~~~~~~~~~~~Set {state:=ABORTED}.}}
\put(0,-6){\makebox(200,95)[l]{~~\textbf{Timer}: If {state=ABORTED} and $T>T_{4}$;}}
\put(0,-18){\makebox(200,95)[l]{~~~~~~~~~~~~~~~Set {ledger} $\mid\mathcal{S}\mid:=${ledger}$\mid\mathcal{S}\mid$+\$Deposit.}}
\put(0,-30){\makebox(200,95)[l]{~~~~~~~~~~~~~~~Set {state:=ABORTED.}}}
\put(0,-10){\makebox(250,20){Alg. 1.  Smart Contract {CS-SevDel}}}
\end{picture}

The verifiability of the data encryption is achieved based on provable data possession and zero-knowledge proofs. The data owners are able to audit the encrypted data by randomly sampling the encrypted blocks. The homomorphic authentication tags guarantee the authentication of data blocks in the aggregated way. First, the homomorphic authentication tags are created in the way of digital signatures. They are not forgeable under the assumption of computational Diffie-Hellman assumption. Second, it is impossible to generate a proof if the cloud server does not encrypt the sampled data blocks because the proof is the linear aggregation of the tags. Therefore, the verifiability of the data encryption is realized.

The erasability of the data is achieved based on the Intel SGX. The enclave is created for the file when the cloud server receives the file. The enclave is used to maintain the decryption key. The deletion of the data is achieved when the enclave is destroyed. The destroy of the enclave would permanently lose the information in the enclave. According to this feature, the decryption key is lost after the destroy of the enclave. Thus, the encrypted file can never be decrypted, so the file is permanently deleted.

The auditability of data leakage is achieved based on the smart contract. The smart contract makes sure the automatic execution of the service-level agreement between the data owners and the cloud service providers. The condition that triggers penalty is the data leakage incident, so the data owner needs to prove to the blockchain node that they have the leaked data. This proof generation method is the same as the method for data encryption proof, so they are based on the same assumption.

\section{Conclusion}\label{sec:con}
In this paper, present a secure and verifiable data deletion scheme that leverages the zero-knowledge proof to achieve the verification of the encryption of the outsourced data without retrieving the ciphertexts. The deletion of the encryption keys are guaranteed based on Intel SGX. The proposed scheme implements secure interfaces to perform data encryption and decryption for secure cloud storage and utilizes smart contract to enforce the operations of the cloud service provider to follow service level agreements with data owners and the penalty over the service provider, who discloses the cloud data on its servers.
As the proposed scheme enables the cloud server to handle the service-side encryption, which make the scheme particularly suitable for the popular secure cloud storage services.

\bibliographystyle{IEEEtran}
\bibliography{Ref}

\begin{thebibliography}{10}
\providecommand{\url}[1]{#1}
\csname url@samestyle\endcsname
\providecommand{\newblock}{\relax}
\providecommand{\bibinfo}[2]{#2}
\providecommand{\BIBentrySTDinterwordspacing}{\spaceskip=0pt\relax}
\providecommand{\BIBentryALTinterwordstretchfactor}{4}
\providecommand{\BIBentryALTinterwordspacing}{\spaceskip=\fontdimen2\font plus
\BIBentryALTinterwordstretchfactor\fontdimen3\font minus
  \fontdimen4\font\relax}
\providecommand{\BIBforeignlanguage}[2]{{%
\expandafter\ifx\csname l@#1\endcsname\relax
\typeout{** WARNING: IEEEtran.bst: No hyphenation pattern has been}%
\typeout{** loaded for the language `#1'. Using the pattern for}%
\typeout{** the default language instead.}%
\else
\language=\csname l@#1\endcsname
\fi
#2}}
\providecommand{\BIBdecl}{\relax}
\BIBdecl

\bibitem{abu2010racs}
H.~Abu-Libdeh, L.~Princehouse, and H.~Weatherspoon, ``Racs: a case for cloud
  storage diversity,'' in \emph{Proceedings of the 1st ACM symposium on Cloud
  computing}, 2010, pp. 229--240.

\bibitem{palankar2008amazon}
M.~R. Palankar, A.~Iamnitchi, M.~Ripeanu, and S.~Garfinkel, ``Amazon s3 for
  science grids: a viable solution?'' in \emph{Proceedings of the 2008
  international workshop on Data-aware distributed computing}, 2008, pp.
  55--64.

\bibitem{quick2014google}
D.~Quick and K.-K.~R. Choo, ``Google drive: Forensic analysis of data
  remnants,'' \emph{Journal of Network and Computer Applications}, vol.~40, pp.
  179--193, 2014.

\bibitem{dropbox2014dropbox}
I.~Dropbox, ``Dropbox,'' \emph{http://www. dropbox. com}, 2014.

\bibitem{holt2018apple}
D.~X. Holt, ``Apple icloud: Securing your data,'' 2018.

\bibitem{wilson2015onedrive}
K.~Wilson and K.~Wilson, ``Onedrive,'' \emph{Everyday Computing with Windows
  8.1}, pp. 71--74, 2015.

\bibitem{kuyoro2011cloud}
S.~Kuyoro, F.~Ibikunle, and O.~Awodele, ``Cloud computing security issues and
  challenges,'' \emph{International Journal of Computer Networks (IJCN)},
  vol.~3, no.~5, pp. 247--255, 2011.

\bibitem{zheng2020toward}
D.~Zheng, L.~Xue, C.~Yu, Y.~Li, and Y.~Yu, ``Toward assured data deletion in
  cloud storage,'' \emph{IEEE Network}, vol.~34, no.~3, pp. 101--107, 2020.

\bibitem{voigt2017eu}
P.~Voigt and A.~Von~dem Bussche, ``The eu general data protection regulation
  (gdpr),'' \emph{A Practical Guide, 1st Ed., Cham: Springer International
  Publishing}, vol.~10, no. 3152676, pp. 10--5555, 2017.

\bibitem{hughes2021symposium}
J.~Hughes, M.~E. Kaminski, J.~Snow, and F.~T. Wu, ``Symposium: The california
  consumer privacy act,'' \emph{Loyola of Los Angeles Law Review}, vol.~54,
  2021.

\bibitem{calzada2022citizens}
I.~Calzada, ``Citizens’ data privacy in china: The state of the art of the
  personal information protection law (pipl),'' \emph{Smart Cities}, vol.~5,
  no.~3, pp. 1129--1150, 2022.

\bibitem{garfinkel2003remembrance}
S.~L. Garfinkel and A.~Shelat, ``Remembrance of data passed: A study of disk
  sanitization practices,'' \emph{IEEE Security \& Privacy}, vol.~1, no.~1, pp.
  17--27, 2003.

\bibitem{liu2017erasucrypto}
C.~Liu, H.~A. Khouzani, and C.~Yang, ``Erasucrypto: A light-weight secure data
  deletion scheme for solid state drives.'' \emph{Proc. Priv. Enhancing
  Technol.}, vol. 2017, no.~1, pp. 132--148, 2017.

\bibitem{gutmann1996secure}
P.~Gutmann, ``Secure deletion of data from magnetic and solid-state memory,''
  in \emph{Proceedings of the Sixth USENIX Security Symposium, San Jose, CA},
  vol.~14, 1996, pp. 77--89.

\bibitem{boneh1996revocable}
D.~Boneh and R.~J. Lipton, ``A revocable backup system.'' in \emph{USENIX
  Security Symposium}, 1996, pp. 91--96.

\bibitem{hao2015deleting}
F.~Hao, D.~Clarke, and A.~F. Zorzo, ``Deleting secret data with public
  verifiability,'' \emph{IEEE Transactions on Dependable and Secure Computing},
  vol.~13, no.~6, pp. 617--629, 2015.

\bibitem{goldreich1996composition}
O.~Goldreich and H.~Krawczyk, ``On the composition of zero-knowledge proof
  systems,'' \emph{SIAM Journal on Computing}, vol.~25, no.~1, pp. 169--192,
  1996.

\bibitem{gotzfried2017cache}
J.~G{\"o}tzfried, M.~Eckert, S.~Schinzel, and T.~M{\"u}ller, ``Cache attacks on
  intel sgx,'' in \emph{Proceedings of the 10th European Workshop on Systems
  Security}, 2017, pp. 1--6.

\bibitem{zheng2018blockchain}
Z.~Zheng, S.~Xie, H.-N. Dai, X.~Chen, and H.~Wang, ``Blockchain challenges and
  opportunities: A survey,'' \emph{International journal of web and grid
  services}, vol.~14, no.~4, pp. 352--375, 2018.

\bibitem{ni2017differentially}
J.~Ni, K.~Zhang, K.~Alharbi, X.~Lin, N.~Zhang, and X.~S. Shen, ``Differentially
  private smart metering with fault tolerance and range-based filtering,''
  \emph{IEEE Transactions on Smart Grid}, vol.~8, no.~5, pp. 2483--2493, 2017.

\bibitem{erway2015dynamic}
C.~C. Erway, A.~K{\"u}p{\c{c}}{\"u}, C.~Papamanthou, and R.~Tamassia, ``Dynamic
  provable data possession,'' \emph{ACM Transactions on Information and System
  Security (TISSEC)}, vol.~17, no.~4, pp. 1--29, 2015.

\bibitem{shacham2013compact}
H.~Shacham and B.~Waters, ``Compact proofs of retrievability,'' \emph{Journal
  of cryptology}, vol.~26, no.~3, pp. 442--483, 2013.

\end{thebibliography}


\end{document}